\documentstyle[12pt]{article}
\begin{document}
\topmargin -1.cm
\baselineskip 0.3in 
\centerline{\large {\bf Micellar Aggregates of Gemini Surfactants:}}
\centerline{\large {\bf Monte Carlo Simulation of a Microscopic Model}} 

\vspace{.5cm} 

\centerline{\bf Prabal K. Maiti$^{1}$ and Debashish Chowdhury$^{1,2,*}$}

\vspace{.25cm}

\centerline{$^1$Department of Physics, Indian Institute of Technology,}
\centerline{ Kanpur 208016, U.P., India$^{\dagger}$}
\centerline{$^2$Institute for Theoretical Physics, University of Cologne}
\centerline{ D-50937 K\"oln, Germany$^{\ddagger}$}
%\begin{document}
%\maketitle
\begin{abstract}
We propose a "microscopic" model of {\it gemini} surfactants in aqueous 
solution. Carrying out extensive Monte Carlo simulations, we study the 
variation of the critical micellar concentration (CMC) of these model 
gemini surfactants with the variation of the (a) length of the spacer 
connecting the two hydrophilic heads, (b) length of the hydrophobic tail 
and (c) the bending rigidity of the hydrocarbon chains forming the spacer 
and the tail; some of the trends of variation are counter-intuitive but 
are in excellent agreement with the available experimental results. 
Our simulations also elucidate the dependence of the shapes of the micellar 
aggregates and the magnitude of the CMC on the geometrical shape and size 
of the surfactant molecules and the electrical charge on the hydrophilic 
heads.  
\end{abstract}
{\bf Running title:} Micellar aggregate of gemini surfactants\\ 
PACS Numbers: 68.10.-m, 82.70.-y 

--------------------------------------------------------------------------

$^{*}$ To whom all correspondence should be addressed;

E-mail: debch@thp.Uni-Koeln.de (till 30th June, 1997). 

$^{\dagger}$ (Permanent Address); E-mail: debch@iitk.ernet.in 

$^{\ddagger}$ (Address till 30th June, 1997) 

\vfill\eject
Soap molecules are common examples of surfactant molecules; these not
only find wide ranging applications in detergent and pharmaceutical 
industries, food technology, petroleum recovery etc. but are also one 
of the most important constituents of cells in living systems. Therefore, 
physics, chemistry, biology and technology meet at the frontier area of 
interdisciplinary research on association colloids formed by surfactants 
\cite{Evans}. 
The "head" part of surfactant molecules consist of a polar or ionic group. 
The "tail" of many surfactants consist of a single hydrocarbon chain 
whereas that of some other surfactants, e.g., phospholipids, are made 
of two hydrocarbon chains both of which are connected to the same head 
\cite{tanford}. In contrast, {\it gemini} surfactants \cite{Deinega,
Menger1,Menger2, Rosen1}, consist of two single-chain surfactants whose 
heads are connected by a "spacer" chain and, hence, these "double-headed" 
surfactants are sometimes also referred to as "dimeric surfactants" 
\cite{Zana1,Zana2}. The gemini surfactants have several unusual properties. 
Some of these properties, which make these very attractive for potential 
industrial use, are crucially influenced by the aggregation of the 
surfactants and the morphologies of these supramolecular aggregates. 
Therefore, in order to gain insight into the physical origin of some of 
the unusual properies of gemini surfactants, in this letter we propose a 
simple microscopic model and study the formation and morphologies of the 
supramolecular aggregates of these model gemini surfactants by Monte Carlo 
(MC) computer simulations. 

When put into an aqueous medium, the "heads" of the surfactants like to get 
immersed in water and, hence, called "hydrophilic" while the tails tend to 
minimize contact with water and, hence, called "hydrophobic" \cite{tanford}. 
The spacer in gemini surfactants is usually hydrophobic but gemini 
surfactants with hydrophilic spacers have also been synthesized\cite{Rosen2}. 
A multi-component fluid mixture containing water and surfactants  
minimizes the free energy by forming "{\it self-assemblies}"  (i.e., 
supra-molecular aggregates) of surfactants, such as monolayer and bilayer 
membranes, micelles, inverted-micelles, vesicles, etc. \cite{Gelbert}.  
Micelles are formed when the concentration of the surfactants in water 
exceeds what is known as the critical micellar concentration (CMC) 
\cite{tanford}. 

On the basis of intuitive physical arguments, it is usually expected that 
a longer hydrocarbon chain should lower the CMC. On the contrary,   
two unusual features of the CMC of gemini surfactants with ionic heads are: 
(i) for a given fixed length of each of the two tails, the CMC 
{\it increases} with the length of the spacer till it reaches a maximum 
beyond which CMC {\it decreases} with further increase of the spacer length  
\cite{Zana1,Zana3, Frindi,goyel}; (ii) for a given length of the spacer, 
the CMC {\it increases} with increasing tail length \cite{Menger1,Menger2}. 
Moreover, the micellar aggregates formed by the gemini surfactants with 
short spacers even at low concentrations just above the CMC are "long, 
thread-like and entangled" \cite{Zana2,Zana4}, in contrast to the spherical 
shapes of the micelles formed by single-chain surfactants at such low 
concentrations. Our aim is to understand the physical origin of these 
unusual properties of gemini surfactants. 

A microscopic lattice model of double-chain surfactants (with a single 
head) in aqueous solution was developed by Bernardes\cite{bern2} by 
modifying the Larson model of single-chain surfactants~\cite{Lar1,St1,Livrev}. 
In this letter we propose a microscopic lattice model of gemini surfactants 
by extending Bernardes' model so as to incorporate two hydrophilic heads  
connected by a {\it hydrophobic spacer}. 

The Larson model was originally developed for ternary microemulsions which 
consist of water, oil and surfactants. In the spirit of lattice gas models, 
the fluid under investigation is modelled as a simple cubic lattice of size 
$L_x \times L_y \times L_z$. Each of the molecules of water (and oil) can 
occupy a single lattice site. A surfactant occupies several lattice sites 
each successive pairs of which are connected by rigid nearest-neighbour bond. 
A single-chain surfactant can be described by the symbol ~\cite{Livrev}
${\cal T}_m{\cal N}_p{\cal H}_q$ where ${\cal T}$ denotes {\it tail}, 
${\cal H}$ denotes {\it head} and ${\cal N}$ denotes the 'liaison' or neutral 
part of the surfactants. $m$, $p$ and $q$ are integers denoting the lengths 
of the tail, neutral region and head, respectively, in the units of lattice 
sites. Thus, each surfactant is a self-avoiding chain of length 
$\ell = (m+p+q)$. The "water-loving" head group is assumed to be "water-like" 
and, similarly, the "oil-loving" tail group is assumed to be "oil-like". 
Bernardes' lattice model of double-chain surfactants with a single 
hydrophilic head can be described by the symbol 
${\cal T}_m{\cal N}_p{\cal H}_q{\cal N}_p{\cal T}_m$. In terms 
of the same symbols, the microscopic lattice model of a gemini surfactant, 
which we propose here, can be represented by the symbol 
${\cal T}_m{\cal N}_p{\cal H}_q{\cal S}_n{\cal H}_q{\cal N}_p{\cal T}_m$ 
where $n$ is the number of lattice sites constituting the spacer represented 
by the symbol ${\cal S}$. We shall refer to each site on the surfactants 
as a $monomer$. 

Jan, Stauffer and collaborators \cite{St1} reformulated the Larson model 
in terms of Ising-like variables, in the same spirit in which a large 
number of simpler lattice models had been formulated earlier \cite{Gomsch} 
for the convenience of calculations. In this reformulation, a classical Ising 
spin variable $S$ is assigned to each lattice site; $S_i = 1$ ($-1$) if the 
$i$-th lattice site is occupied by a water (oil) molecule. If the $j$-th 
site is occupied by a monomer belonging to a surfactant then $S_j = 1, -1, 0$ 
depending on whether the monomer at the $j$th site belongs to head, tail or  
neutral part. The monomer-monomer interactions are taken into account through 
the interaction between the corresponding pair of Ising spins which is assumed 
to be non-zero provided the spins are located on the nearest-neighbour sites 
on the lattice. Thus, the Hamiltonian for the system is given by the 
standard form 
\begin{equation}
H = - J \sum_{<ij>} S_i S_j. 
\end{equation}
where attractive interaction (analogue of the ferromagnetic interaction in 
Ising magnets) corresponds to $J > 0$ and repulsive interaction (analogue 
of antiferromagnetic interaction) corresponds to $J < 0$ \cite{St1}. 
Temperature $T$ of the system is measured in the units of $J$ (the Boltzmann 
constant $k_B = 1.0$). 

We have considered three possible Larson-type microscopic lattice models 
of $ionic$ gemini surfactants. In the simplest model, which we call model 
$A$, the monomers belonging to heads have Ising spin $+2$ to mimic the 
presence of charge. The repulsive interaction between a pair of ionic 
heads is taken into account through an antiferromagnetic interaction 
$J = -1$ between pairs of nearest neighbour sites both of which carry 
spins $+2$; however, the interaction between all other pairs of 
nearest-neighbour spins is assumed to be $J = 1$. The short-range of 
the repulsive (antiferromagnetic) interaction between the "charged" 
heads corresponds to very strong screening of the Coulomb repulsion 
between ionic heads by the counterions. Molecular dynamics (MD) 
simulations of a similar molecular model of gemini surfactants has been 
carried out by Karaborni et al. \cite{Karaborni}. In this letter we 
summarize only the most important results on the model $A$ with $hydrophobic$ 
spacer; the results for the models $B$ and $C$ will be reported, together 
with the results for the model $A$ with hydrophilic spacer, in a longer 
paper elsewhere \cite{Maiti}.  

In order to investigate the influence of the ionic heads on the results,   
we have also considered a model of gemini surfactants with non-ionic 
polar heads which is obtained from the model $A$ by replacing 
all the $+2$ Ising spin variables by Ising spin $+1$ (and, accordingly, 
the interactions $-1$ between the heads on nearest-neighbour sites are 
replaced by $+1$). Moreover, in order to investigate the role of the 
chain stiffness we have introduced a chain bending energy; every bend of 
a tail or a spacer, by a right angle at a lattice site, is assumed to 
cost an extra amount of energy $K (>0)$. 

We have carried out MC simulations of the model 
${\cal T}_m{\cal N}_p{\cal H}_q{\cal S}_n{\cal H}_q{\cal N}_p{\cal T}_m$ 
of gemini surfactants for $p = q =1$ and for three differnet values of the 
tail length, namely, $m = 5, 15$ and $25$ in water where 
$L_x = L_y = L_z = 100$. The moves allowed for the surfactants in our 
model are same as described in ref.\cite{Livrev}. In reality, CMC is not 
a single concentration (perhaps, it is more appropriate to call it 
characteristic micellar concentration~\cite{St1}). Following Stauffer et 
al.\cite{St1}, we identify CMC as the amphiphile concentration where half 
of the surfactants are in the form of isolated chains and the other half 
in the form of clusters consisting of more than one neighbouring amphiphile.    
For a given $m$ we have computed the CMC for spacer lengths $2 \leq n \leq 20$. 

The non-monotonic variation of CMC of ionic gemini surfactants with the 
spacer length, shown in figs. 1 and 2, is in qualitative agreement with the 
experimental observations \cite{Zana3,Frindi,goyel,Zana4}. Moreover, for a 
given length of the spacer, the CMC increases when the bending stiffness $K$ 
of the hydrophobic chains is switched on. Furthermore, we have observed that, 
for a given length of the hydrophobic spacer, the CMC of ionic gemini 
surfactants {\sl increase} with the increase of the tail length~\cite{Maiti}; 
this trend of variation is also consistent with the corresponding experimental 
observations \cite{Menger1,Menger2}. 

For a given tail length, the CMC of model gemini surfactants with non-ionic 
polar head groups decreases $monotonically$ with the increase in the spacer 
length for both $m=5$ and $m=15$ (see fig.3). This is in sharp contrast to 
the non-monotonic variation observed for ionic gemini surfactants. However, 
for a given spacer length, the trend of the variation of CMC of non-ionic 
gemini surfactants with the tail length is similar to that observed for ionic 
gemini surfactants. 

The snapshots of the micellar aggregates formed by the gemini surfactants 
with ionic heads are shown for spacer length $n = 2$ (fig. 4) and for 
$n = 16$ (fig.5). The morphology of the aggregates in fig.4 are similar to 
the "long, thread-like and entangled" micelles observed in laboratory 
experiments \cite{Zana2} and in MD simulations \cite{Karaborni} on gemini 
surfactants with short spacers. Moreover, our data in fig.5 suggest that 
rod-like micelles are formed by gemini surfactants with $m=15$ when 
the spacer length is $n = 16$. The morphologies of the aggreagtes in 
fig.4 and 5 are in sharp contrast with the spherical shape of the micelles 
(see fig.6) formed by single-chain ionic surfactants of comparable tail 
size even at concentrations somewhat higher than those in the figures 4 and 5.  

We did not observe any significant difference in the shapes of the 
aggregates of ionic and non-ionic gemini surfactants for given values 
of $m$, $n$ and comparable concentration~\cite{Maiti}, in spite of 
qualitatively different trends of variation of CMC with spacer lengths. 

Therefore, we conclude that (i) the shapes of aggregates are dominantly 
determined by the geometric shape and size of the molecules whereas 
(ii) the variation of CMC with spacer length is strongly influenced by the 
ionic charge. It would be interesting to investigate the effects of 
weakening of the screening (i.e., increasing the range) of the repulsive 
Coulomb interaction between the ionic heads on the results reported in this 
letter; but, such a MC study will require much larger computational resources. 

{\bf Acknowledgements:} One of us (DC) thanks V.K. Aswal, A.T. Bernardes, 
S. Bhattacharya, P.S. Goyal, D. Stauffer and R. Zana for useful 
discussions/correspondences and the Alexander von Humboldt Foundation for 
partial financial support.

\newpage 

\noindent{\bf Figure Captions:} 

\noindent{\bf Fig.1:} Variation of CMC of ionic geminis  
with spacer length; $m = 15$, $T = 2.2$. The symbols $\Box$ and $\times$ 
correspond to $K = 0$ and $K = 2$, respectively. The continuous 
curves are merely guides to the eye. 

\noindent{\bf Fig.2:} Same as fig.1, except that $m = 5$. The symbols 
$\triangle$ and $\ast$ correspond to $K = 0$ and $K = 2$, respectively.  

\noindent{\bf Fig.3:} Variation of CMC of non-ionic geminis with spacer  
length; $m = 15$ ($\Box$) and $m = 5$ ($\triangle$) both with $K = 0$ 
and at $T = 2.2$. The continuous curves are merely guides to the eye.  

\noindent{\bf Fig.4:} Snapshots of the micellar aggregates formed by 
ionic geminis with $m = 15$, $n = 2$ and $K = 0$ at $T = 2.2$ when  
the surfactant density is $0.007$. 
The symbols black spheres, dark grey spheres and light grey spheres 
represent monomers belonging to head, tail and spacer, respectively. 

\noindent{\bf Fig.5:} Same as in fig.5, except that $n = 16$ and the 
density is $0.005$. 

\noindent{\bf Fig.6:} Snapshots of micellar aggregates formed by 
single-chain ionic surfactants with $m = 14$ and the density $0.01$.
The symbols black spheres and grey spheres represent monomers 
belonging to head and tail, respectively. 


\newpage 
\begin{thebibliography}{99}
\bibitem{Evans}  Evans D. F. and Wennerstrom H., {\sl The colloidal
domain where physics, chemistry, biology and technology meet},
VCH, New York (1994).
\bibitem{tanford} Tanford C.,
{\sl The Hydrophobic Effect: Formation of Micelles and Biological
Membranes}, (Wiley, New York 1980).
\bibitem{Deinega} Deinega Y. F., Ulberg Z. R., Marochko L. G., Rudi V. P. and 
Deisenko V. P., Kolloidn Zh. {\bf 36}, 649 (1974).
\bibitem{Menger1} Menger F. M. and Littau C. A., J. Am. Chem. Soc.,
{\bf113}, 1451 (1991).
\bibitem{Menger2} Menger F. M. and Littau C. A., J. Am. Chem. Soc., 
{\bf 115}, 10083 (1993).
\bibitem{Rosen1} Rosen M., Chemtech 30 (1993).  
\bibitem{Zana1} Zana R., Benrraou M. and Rueff R., Langmuir, {\bf7}
,1072 (1991).
\bibitem{Zana2} Zana R. and Talmon Y., Nature, {\bf 362}, 228 (1993).
\bibitem{Rosen2} Song L. D. and Rosen M. J., Langmuir, {\bf 12}, 1149 (1996).
\bibitem{Gelbert} Gelbert W., Ben Shaul A. and Roux D. (eds.) {\sl Micelles, Membranes, Microemulsions and Monolayers} (Springer, Berlin, 1994). 
\bibitem{Zana3} Alami E., Beinert G., Marie P. and Zana R., Langmuir, {\bf 9}, 1465 (1993).
\bibitem{Frindi} Frindi M., Michels B., Levy H. and Zana R., Langmuir, {\bf 10}, 1140 (1994).
\bibitem{goyel} De S., Aswal V. K., Goyal P.S. and Bhattacharya S.,
J. Phys. Chem. {\bf 100}, 11664 (1996). 
\bibitem{Zana4} Hirata H., Hattori N., Ishida M., Okabayashi H., Frusaka M. and Zana R., J. Phys.Chem. {\bf 99}, 17778 (1995). 
\bibitem{bern2} Bernardes A. T., J. Phys. II France {\bf 6}, 169 (1996); 
see also Langmuir, {\bf 12}, 5763 (1996).  
\bibitem{Lar1}
Larson R. G., Scriven L. E. and Davis H. T., J. Chem. Phys. {\bf83}, 2411 
 (1985); Larson R. G., J. Chem. Phys. {\bf89}, 1642 (1988) and 
{\bf 91}, 2479 (1989). 
\bibitem{St1} Stauffer D., Jan N., He Y., Pandey R. B., Marangoni D. G. and 
       Smith-Palmer T., J. Chem. Phys. {\bf 100}, 6934 (1994);  
Jan N. and Stauffer D., J. de Physique I {4}, 345 (1994).  
\bibitem{Livrev}
  Liverpool T. B., in: {\sl Annual Reviews  
  of Computational Physics}, vol. IV, ed. D. Stauffer (World
  Scientific, Singapore 1996). 
\bibitem{Gomsch} Gompper G. and Schick M.,
{\sl Phase Transitions and Critical Phenomena}, Vol. 16,
ed. C. Domb and J. L. Lebowitz, (Academic Press, London 1994).
\bibitem{Karaborni} Karaboni S., Esselink K., Hilbers P. A. J., Smit B.,
 Karthauser J., van Os N. M. and Zana R., Science, {\bf266}, 254 (1994)
\bibitem{Maiti} Maiti P. K. and Chowdhury D., to be published 

\end{thebibliography}
\end{document}